\documentclass{elsart}

\usepackage{amssymb}
\usepackage{graphicx}
\begin{document}

\begin{frontmatter}

\title{A Tiling Approach to Counting Inherent Structures in Hard Potential Systems}% Force line breaks with \\
\author{ S. S. Ashwin}
\address{Department of Chemistry, University of Saskatchewan,
Saskatoon, Saskatchewan, S7N 5C9}
\author{R. K. Bowles}
\address{Department of Chemistry, University of Saskatchewan, 
Saskatoon, Saskatchewan, S7N 5C9}

%\date{\today}% It is always \today, today,
             %  but any date may be explicitly specified
\begin{abstract}
The number of distinguishable inherent structures of a liquid is the key component to understanding the thermodynamics of glass formers. In the case of hard potential systems such as hard  discs, spheres and ellipsoids, an inherent structure corresponds to a collectively jammed configuration. This work develops a tiling based approach to counting inherent structures that constructs packings by combining sets of elementary locally jammed structures but eliminates those final packings that either, do not tile space, or are not collectively jammed, through the use of tile incompatibility rules. The resulting theory contains a number of geometric quantities, such as the number of growth sites on a tile and the number of tile compatibilities that provide insight into the number of inherent structures in certain limits. We also show that these geometric quantities become quite simple in a system of highly confined hard discs.
 \end{abstract}

\end{frontmatter}
%\pacs{64.70Pf,  61.43Fs,}
%\maketitle

% introduce general concept of ideal glass
\section {Introduction}
\label{sec:intro}

% inherent structures
 The connection between the thermodynamics and the way particles pack in dense or supercooled liquids was introduced by Stillinger and Weber~\cite{SW1} through their inherent structure mapping formalism. In this approach, every point on the potential energy landscape (PEL)~\cite{Deb1}, which describes the high-dimensional $N$-body configurational potential energy function, is uniquely mapped to its nearest energy minimum or the mechanically stable, or jammed, configuration. The structure corresponding to this jammed configuration is known as an inherent structure. By grouping together configurations that map to the same inherent structure into local basins of attraction, it is possible to express  the partition function as a sum over the inherent structures and their associated local basin. This gives rise to two distinct entropies: the {\it configurational entropy} associated with the number of accessible inherent structures and the {\it vibrational entropy} due to the degrees of freedom within the basin. Adam and Gibbs~\cite{AG} suggested that the observed slowing down of the dynamics in supercooled liquids with cooling resulted in a dramatic decrease in the number of accessible basins on the potential energy landscape. In principle, if the liquid could be cooled sufficiently, without either crystallizing or getting kinetically trapped in higher potential energy minima, it might eventually become trapped in the single deepest liquid basin, which would represent an ideal glass.
  
% hard sphere
In the case of hard potential systems, one partitions the density landscape (DL)~\cite{DLS} by mapping each configuration of the fluid at a density $\rho$ to a its closest density maximum, $\rho_0$, which is the inherent structure and corresponds to a collectively jammed packed configuration~\cite{cjam}. The number of inherent structures for a hard potential system is expected to be given by~\cite{Deb1},
\begin{equation}
\Omega(\rho_0)=\exp[N s_c(\rho_0)]\mbox{ ,}\\
\label{ng}
\end{equation}
where $N$ is the number of particles and $s_c$ is the configurational entropy per particle. The ideal glass transition would correspond to a density $\rho$ when the liquid becomes trapped in a single basin so that  $\Omega(\rho_{0})$ becomes sub-exponential {\it i.e.,} when $s_c=0$. 

While packing algorithms can exhaustively generate all the inherent structures in small systems~\cite{CircPack1,Szabo,5disks,ohern}, an exact enumeration of packings for thermodynamically sized systems is only available for very simple models~\cite{rich2}.  Since it is difficult to create and count random packings, another approach is to focus on counting the jammed structures that can be generated from lattice structures ~\cite{torq1}. Computer simulations of binary hard disc mixtures~\cite{speedyD} that estimate $\Omega(\rho_{0})$ by comparing the entropy of the equilibrium liquid to that of the system constrained to the basin of one of the inherent structures to which the liquid maps, suggest that the distribution of inherent structures is Gaussian. However, it is difficult to maintain equilibrium in the fluid at high densities and the ideal glass density is obtained by extrapolating the distribution to higher densities. The existence of the ideal glass clearly depends on the details of this extrapolation~\cite{1dglass,deb03}. Simulations of polydispersed hard spheres~\cite{Warner} at high densities point to an absence of a thermodynamic glass transition and a recent simulation of a binary mixture of hard discs~\cite{Donev} finds an exponential number of inherent structures up to the crystallization density, which suggests there is no ideal glass transition in this system.

%stillinger tiles and bounds.
A physically motivated argument~\cite{Still2} shows that the number of {inherent structures} for systems with short range potentials is always exponential with respect to the system size.  However, a key element of this approach requires the division of volume into molecularly sized, but mechanically stable, subvolumes. This ignores the more global nature of the unjamming process in hard potential systems~\cite{cjam} which may lead to a reduction in the number of jammed structures. The goal of the present paper is to develop a tiling approach to counting inherent structures in systems with hard core potentials that will account for the collective nature of unjamming. Section~\ref{sec:tiles} introduces the details of the tiles and tile incompatibilities while Section~\ref{sec:bound} examines some of the properties of the geometric quantities arising from our method. The method is applied to a system of confined discs in Section~\ref{sec:2d} and Section~\ref{sec:disc} contains our discussion and conclusions.

% sections description - Can be removed if submit a letter.
%Our argument is developed in the context of a tiling model and we begin in Section II with our definitions of tiles and tile incompatibility. Section III describes the enumeration of inherent structures and our results are discussed in Section IV. 

\section{Tiles and tile incompatibility}
\label{sec:tiles}
Locally jammed hard particles are unable to move, due to the nonoverlap requirement of the potential, if the neighboring particles in contact are held fixed~\cite{cjam}. For example, a two dimensional hard disc will be jammed if it has at least three fixed contacts that are not all in the same semicircle. A packing constructed entirely of locally jammed  particles may not  be collectively jammed because collective motions of several particles could allow the packing to collapse. However, a collectively jammed state must consist entirely of locally jammed particles.  This suggests that we may be able to construct collectively jammed configurations by adding locally jammed particles to a ``growing" packing, in a manner similar to Bernal's original work on liquid structure \cite{bern}, but after each addition, we eliminate those configurations that would result in packings that are not collectively jammed.

%The first step in our process requires a mathematical construction that maps collectively jammed particle packings into a tiling such that we can identify the set of what we will refer to as first order or elementary tiles. This remains a significant theoretical challenge which we do not attempt to resolve here. 
Here, we explore the possibility of counting inherent structures by using this configurational growth process and representing local packing environments of the particles as tiles. Several possible methods for defining these tiles exist, including the Voronoi and Delaunay tesselations~\cite{vor} and the construction by Blumenfield and Edwards (BE)~ \cite{SamEd2,SamEd} that appears to provide a description of the phase space of granular and cellular systems by tiling the plane (in two dimensions) with quadrilaterals referred to as quadrons.  Most reasonable methods of mapping jammed particle configurations to tiles will result in polygon shaped objects  that  can be combined with a set of rules that ensure the local packing conditions are satisfied. For example, in the simplest case, edge matching rules will be required to guarantee local space filling. Additional global restrictions will be required to ensure that $N$ tiles cover all of the volume $V$ without leaving gaps ({\it tiling condition}), and that the arrangement is collectively jammed. Since the addition of each set of constraints for local jamming and collective jamming reduces the number of allowable states at a given density, it follows that 
\begin{equation}
\Omega(\rho_0)\leq\Omega_{l}(\rho_0)\leq\Omega_{t}(\rho_0)\mbox{,}\\
\end{equation}
where $\Omega_{l}(\rho_0)$ and $\Omega_{t}(\rho_0)$ are the number of configurations with density $\rho_0$ that are both locally jammed and satisfy the tiling condition, and those that satisfy the tiling condition alone, respectively. There has been a considerable amount of work on packing algorithms in confined systems ~\cite{Szabo} and these algorithms can be used to construct a large enough configuration from which one can identify a set of tiles that represent all the possible local jamming environments. 

We begin by defining the first order {\it prototile set}  $T_1=\{t^1_1,t^2_1...\}$  of the tilings which give rise to jammed configurations of  density $\rho_{0}$. The following should hold for the first order prototiles \cite{tile} (from now on, referred to as first order tiles): (a) tiles are (closed) topological discs, (b) the tiles  are mutually non-congruent and (c) every tile should enclose one particle.  These conditions should allow every inherent structure to be constructed from the elements of $T_1$. An underlying assumption in our scheme is that, for jammed configurations, one is restricted to a countable number of local arrangements and particle neighborhoods, giving rise to a finite set of non-congruent tiles. In a system with periodic boundaries in all directions, one might expect the tiles to have an unrestricted number of allowable orientations. By imposing linear boundary conditions, the rotational degrees of freedom of the tiles should be restricted to a finite set of rigid motions. 

Each element of $T_1$, $t^i_1$, where the superscript $i$ identifies a unique member of the set and the subscript 1 denotes the fact that these tiles are elementary or first order tiles, represents a single particle and its packing environment. A tile consists of a set of boundary points, $b^i_1$, that denotes the edge of the tile, and a set of interior points $v^i_1$.  An arrangement of $N$ tiles chosen from $T_1$,  contained in the area or volume to be tiled, $H$, is called a {\it valid tiling} of $H$, if and only if: (i) {\it no gap exists between the tiles i.e.,} all points $x\in H$ belong to at least one tile and every point in the tiles belongs to $H$,  (ii) {\it there is no overlapping of tiles i.e.,} for every pair of first order tiles, $t^i_1$ and $t^j_1$, in the tiling, $v^i_1\cup v^j_1=\emptyset$ and (iii) {\it there is collective jamming i.e., } the configuration arising out of the tiling of $H$ should be collectively jammed. If these criteria do not hold for an arrangement of tiles, it is called an {\it invalid tiling}.  This defines our {\it tiling} requirement. Every inherent structure of density $\rho_{0}$ is  a valid tiling of $H$.  

Two tiles, $t^i_1$ and $t^j_1$, are {\it neighbors} if they do not overlap but share boundaries.  If two tiles $t^i_1,t^j_1\in T_1$ are neighbors,  the resulting tile $t^l_2=t^i_1\cup t^j_1$ is said to be a tile of {\it order two} if they match edges. The elements of the set $T_2$ should satisfy (a),(b) and  should enclose two particles. For tiles of order two or more,
the criterion (a) is qualified further, {\it i.e.,} two tiles need not be mutually congruent if the arrangement of the particles in the tiles is different.  In general we define a $k^{th}$ {\it order tile} as the union of  a tile $t^i_1$ with a {\it neighbor} tile of order $k-1$.  The elements of $T_k$  should satisfy (a),(b) and enclose $k$ particles. In addition, it is critical to test any $k^{th}$ order tile for  jamming at that level. This can be achieved by fixing each of the particles in the boundary tiles and randomly perturbing the interior particles to test for unjamming motions~\cite{Donev2}. If the $k^{th}$ order tile turns out to be globally jammed at its level, it is said to belong to $T_k$ and we  progress to the $k+1^{th}$ order.  This describes how we create higher order tiles by successively adding first order tiles to the growing structure.

We now introduce incompatibility rules, along with a number of other parameters, to capture the effects of geometric frustration in the inherent structure enumeration.  Let $n_1^{ij}$ be the number of distinct ways we can edge match tiles $t_1^i$ and $t_1^j$ as neighbors, such that they belong to $T_2$. This also represents the number of sites where we can join the two tiles. If $n_1^{ij}=0$ for a particular pair of tiles, we say these tiles are incompatible. The total number of incompatibilities of $t_1^i$ with the other tiles of $T_1$ is the degree of incompatibility, $c^i_1$.
The notion of incompatibility can be generalised to higher order tiles if $n^{ij}_{k}$ is defined as the number of distinct ways we can add $t_1^j$ to the $k^{th}$ order tile, $t^i_{k}$, to generate a new $k+1$ order tile belonging to $T_{k+1}$. Accordingly, the degree of incompatibility  $c_k^i$,  is the number of first order tiles $t_k^i$ is incompatible with.

If $P_1$ is the number of first order tiles, then the number of second order tiles is given by
 \begin{equation}
 P_2 = f_2\sum_{j=1}^{P_1} \sum_{i=1}^{P_{1}}n^{ji}_1 \label{eq_p2}\mbox{ ,}
 \end{equation} 
where $f_2$ is the fraction of the newly generated tiles that are mutually non-congruent in general (unless the configuration of particles in the congruent tiles is different). Similarly, one defines $f_3,f_4...,f_N$ and $n^{ij}_2,n^{ij}_3....,n^{ij}_N$ for higher order tiles. The fractions $\{f_i\}$ depend upon the geometric symmetry of the tiles and will depend on the kind of mapping procedure employed to map configurations to tiles {\it e.g.,} $f_i$ may be different when the tiles are obtained from a Voronoi construction rather than a quadron construction~\cite{SamEd,SamEd2}. Extending this argument, one can write the number of ways of obtaining distinct $k^{th}$ order tiles as
\begin{equation}
P_k = f_k\sum_{j=1}^{P_1} \sum_{i=1}^{P_{k}}n^{ji}_{k-1} \label{eq_p2a}\mbox{ .}
\end{equation} 
While this approach is quite general, it is apparent that $f_{k}$ and $n^{ji}_{k-1}$ are going to be complicated functions of $k$ that may be difficult to obtain for many models. For example $\{f_i\}$ will depend upon the geometric symmetry of the tiles, but will lie in the range $1\geq f_k\geq \frac{1}{P_k}$, while $n^{ji}_{k-1}$ will be dependent on the surface area of a given $k^{th}$ order tile. However, as we show in section 4, these quantities become quite simple if we consider highly confined packings, so this approach may be useful under these circumstances.
 
 \section{Bounds on enumeration}
 \label{sec:bound}
The enumeration of inherent structures is at the heart of the problem of the ideal glass transition.  Stillinger \cite{Still2} argues that the upper and lower limits of the number of distinguishable inherent structures is exponential in system size. The lower bound becoming  exponential in system size  precludes the formation of an ideal glass. The main assumption in this argument is that the system can be divided into subvolumes which are large enough that the defects interaction between neighboring subvolumes is weak. This divides the system into independent and permutationally equivalent regions, the combinations of which correspond to distinguishable inherent structures. In our methodology, the tiles are not permutationally equivalent and the compatibility rules between the neighboring tiles govern whether a configuration is an inherent structure or not.   

We calculate  bounds to the number of packings obtained by taking different limiting cases on the degree of compatibility. We expect $n^{ij}_k$ to be proportional to the surface area of the $k^{th}$ order tile and the maximum value to occur in tiles that are highly ramified.  $n^*_k$ should initially increase with increasing tile order, but must eventually go through a maximum and start to decrease as the tile percolates through space and the addition of new first order tiles begins to fill in the gaps. Given the complexity of  $n^{ji}_{k-1}$, we simply define its maximum value $n^*_k=max\{\sum_{i}n^{ji}_k\}$. Following on from Eq.~\ref{eq_p2}, the upper limit on the number of $k^{th}$ order tiles is given by 
 \begin{equation}
P_{k} \leq P^{*}_{k} = f_k n^*_{k-1}\sum_{j=1}^{P^{*}_{k-1}}P_1- c^j_{k-1}\mbox{, }\\
 \label{eq_pm}
 \end{equation}
which provides the upper bound to the number of inherent structures we can generate as 
 \begin{equation}
 \Omega(\rho_0)\le P^{*}_N =  f_N n^*_{N-1} \sum_{j=1}^{P^{*}_{N-1}}P_1-c^{j}_{N-1}\mbox{. }\\
 \label{eq_pn}
 \end{equation}
If we choose $c^j_m=0$, for all $j,m$, we obtain the  maximum value $P^{*}_N$ as
\begin{equation}
(P^{*}_N)_{max}=\prod_{k=2}^{N}n_{k-1} f_k P_1^N\mbox{, }
\label{eq_pnmax}
\end{equation}
which shows $\Omega(\rho_0)$ increases exponentially with $N$. This is  consistent with Sillinger's upper bound ~\cite{Still2}.
 
 % This is  a much stronger condition for the existence of an ideal glass than required. [need to expand on this in discussion]
It is interesting to see if our relations will allow for a subexponential enumeration of inherent structures. This is possible 
when the number of new tiles generated from an existing tile is minimized by maximizing the number of incompatibilities so that, for any given tile, only one of the $P_1$ first order tiles can be added. This leads to the constraint, $c_{k-1}^j=P_1-1$ for all $j,k$, and gives rise to the following recursive series: 
\begin{eqnarray}
 (P^{*}_{2})_{min}&=&\sum_{j}^{P_1} n_1 f_2=n_1 f_2 P_1\\\nonumber
  (P^{*}_{i})_{min}&=&\sum_{j=1}^{(P^{*}_{i-1})_{min}}n_i f_i=\prod_{k=2}^{i}n_{k-1}f_{k}P_1\\\nonumber
(P^{*}_N)_{min}&=&\prod_{k=2}^{N}n_{k-1}  f_k P_1\mbox{. }
\end{eqnarray}
Since there can only be one inherent structure, $\{f_i\}$ and $\{n_i\}$ must satisfy
\begin{equation}
\prod_{k=2}^{N}n_{k-1} f_k\sim\frac{1}{P_1}\mbox{. }
\label{eq_subex}
\end{equation}
In a simple crystal with only one tile type, $P_1=1$, so the upper bound given by Eq.~\ref{eq_pnmax} is equal to the lower bound given by Eq.~\ref{eq_subex} and $\prod_{k=2}^{N}n_{k-1} f_k=1$. In the case of an ideal glass, we might expect $P_1>1$, but we also note the constraint that only one new tile is generated at each $k$ is much stronger than would be required for a system to have only one inherent structure. We would like to once again emphasize that, since the boundary is fixed, the number of possible orientations of a tile is finite. Stillinger's subvolumes implicitly invoke the same assumptions.

\section{Packings of highly confined discs in two dimensions}
\label{sec:2d}
 In general, $n_k$ and $f_k$ will be complicated functions but they become simplified in confined systems. As an example, we will consider a simple system of hard discs of diameter $\sigma$, confined between two hard lines separated by a distance $1<H/\sigma<1+\sqrt{3/4}$, for which the complete enumeration of the inherent structures was investigated earlier \cite{rich2}.  Due to the narrow separation of the walls, the discs are only able to contact one nearest neighbour on each side and the walls. This only allows for two distinct conformations: (a) two adjacent discs have the same height in the channel (for example tile 3-4 in Fig.~1) or (b) when they contact across the channel (e.g. tiles 2-3, 4-5 in Fig.~1). One readily obtains a binomial distribution for the enumeration of the inherent structures with respect to  $\rho_o$. Our present implementation for the counting is more cumbersome than necessary for this model, but it provides a useful comparison and helps to illustrate our scheme which is more generally applicable. Additionally, the model has the advantage that any $k^{th}$ order tile is naturally jammed, hence no additional perturbation check is required. Fig.~1 shows two distinct configurations which are enough to illustrate the tiles that can be used to construct any jammed state for this system. The limited number of geometrical arrangements for the local packings makes the first order tile set easy to obtain (see Fig.~1). A total of 14 distinct tiles can be identified, numbered from 1 to 7, and their mirror images parallel to the walls, which are primed. An arrangement of tiles (2-3-4-5) is referred to as a ``defect". The top configuration in Fig.~1 shows an isolated defect while the bottom shows two neighboring defects. These configurations give rise to different tiles due to the way space is shared between the discs.

We are now in a position to construct the inherent structures. Building the tiles from left to right such that the right tile disc should touch the left disc and using edge matching for the compatibility rules yields 30 second order tiles : 1-1',1-2',2-3,3-4,3-4',3-6,3-6',4-5,5-1',5-2',6-7',7-6',7-6,7-4,7-4',  1'-1,1'-2,2'-3',3'-4',3'-4,3'-6',3'-6,4'-5',5'-1,5'-2,6'-7,7'-6,7'-6',7'-4',7'-4. Note that  6-7' and 6'-7' are of the same shape, but yield different confirmation of particle positions and are still distinct. Therefore, we have 30 distinct tiles and hence $f_2=1$. Using these 30 tiles, we generate the following third order tiles:
1-1'-1,1-1'-2,1-2'-3',2-3-4,2-3-4',2-3-6,2-3-6',3-4-5,3-4'-5',3-6-7',3-6'-7',4-5-1',4-5-2',5-1'-1,5-1'-2,5-2'-3',6-7'-6',6-7'-6,6-7'-4',6-7'-4,7-6'-7',7-6-7',7-4-5,7-4'-5'. We can complement this sequence of  tiles ({\it e.g., A-B'-C' $\rightarrow$ A'-B-C}) and get their corresponding mirror images, which are also valid tiles of order three. Continuing this process would give rise to all possible jammed configurations.  In Fig.~2, we compare the number of inherent structures obtained from our tile counting mechanism and the actual number of inherent structures for this model. The enumeration by tiling is larger by orders of magnitude because the mapping process produces different tiles capable of producing the same start and end to a configuration. This problem of over counting can be eliminated by storing the distinctly different inherent structures obtained and calculating the Euclidean distance between the new structure obtained from the tiling at the last stage. If a configuration is degenerate, it is rejected.

\section{Discussion}
\label{sec:disc}
The main goal of the present work was to develop an approach for counting the inherent structures of hard core potential models such as hard discs and hard spheres. A key element of our approach is to note that the collectively jammed states, or inherent structures, must be constructed by combining sub-units of locally jammed states. Recent simulations studies of hard sphere packings~\cite{aste06,anik07} have pointed to the important role of tetrahedral packing units.  Anikeenko and Medvedev~\cite{anik07} show that the volume fraction of polytetrahedra increases as the density of the packing increases, with the most dense amorphous packing occurring when all $N$ spheres of the packing are involved in the tetrahedral sub-units. These elementary structures could be represented by the first order tiles in our approach. However, building a packing from locally jammed structures does not ensure that it will tile space, or, even if it does, that the packing will be collectively jammed. Our method deals with these geometric constraints by introducing tile incompatibility.  In particular, our incompatibilities are tested by checking if a  $k^{th}$ tile can combine with a first order tile in order to collectively jam, using an appropriate test (such as perturbation) of  the collectively jammed state.  Our criteria for an incompatibility accounts for the global nature of jamming even though the tile construction is based on local geometric factors. This sets our approach apart from earlier analytical approaches and is important because unjamming events can involve a significant proportion of the particles in the system.

Our tiling approach requires a knowledge of all the local packing environments, but this is usually not known {\it a priori}.
 To make our approach truly useful, we suggest the following practical implementation: We quench a few configurations of a large system, or use local packing algorithms \cite{CircPack1,Szabo} for densities which are not obtainable from quenches, to generate a set of collectively jammed states. Then, we use this set of configurations to identify both the local structures that correspond to the first order tiles and the incompatibilities. It should be possible to identify the local packing structures by studying a subset of the inherent structures, at a given density, using large system sizes. The main challenge is to choose order parameters that clearly distinguish each of the tiles but, as discussed above, progress is already being made in this direction. Some incompatibilities can also be identified by studying collectively jammed states generated by simulation and recognising which tile combinations occur, and which ones do not. Alternatively, algorithms that identify the unjamming motions of a packing could be used to identify those combinations of locally jammed structures that are not collectively jammed~\cite{Donev2}.

One of the authors, SSA, would like to thank Niels Ellegaard for his comments on the manuscript.
Support for this work was provided by NSERC and CFI.

\newpage
{\bf Figure CaptionS}\\
{\bf Fig. 1:}  Tiles shown can be used to construct a jammed state for a model when discs of diameter $\sigma$ are  held within two hard walls separated by a height $H$, such that $H/\sigma < 1.86$. The tiles here are numbered from 1-7' (not all shown). The primed and the unprimed tiles have a reflection symmetry
 along the axis parallel to the wall, dividing the system into halves. Two typical configurations are shown, the top has one ``defect" (tile combination 2-3-4-5) and the bottom has two neighboring defects (tile combination (2'-3'-6'-7'). The fourteen tiles (1-7') can be used to construct any jammed configuration
 for the above model.\\
 {\bf Fig. 2 :} Inherent structure enumeration obtained via the present tiling method (circles) and the actual enumeration for a simple model
  described in the text \cite{rich2} (diamonds).\\ \\
 
 \newpage
 \newpage
 
%%%%%%%%%%%%%%%%%%%%%%%%%%%%%%%%%%%
\begin{figure}[b]
\centering
 \includegraphics[width=185mm]{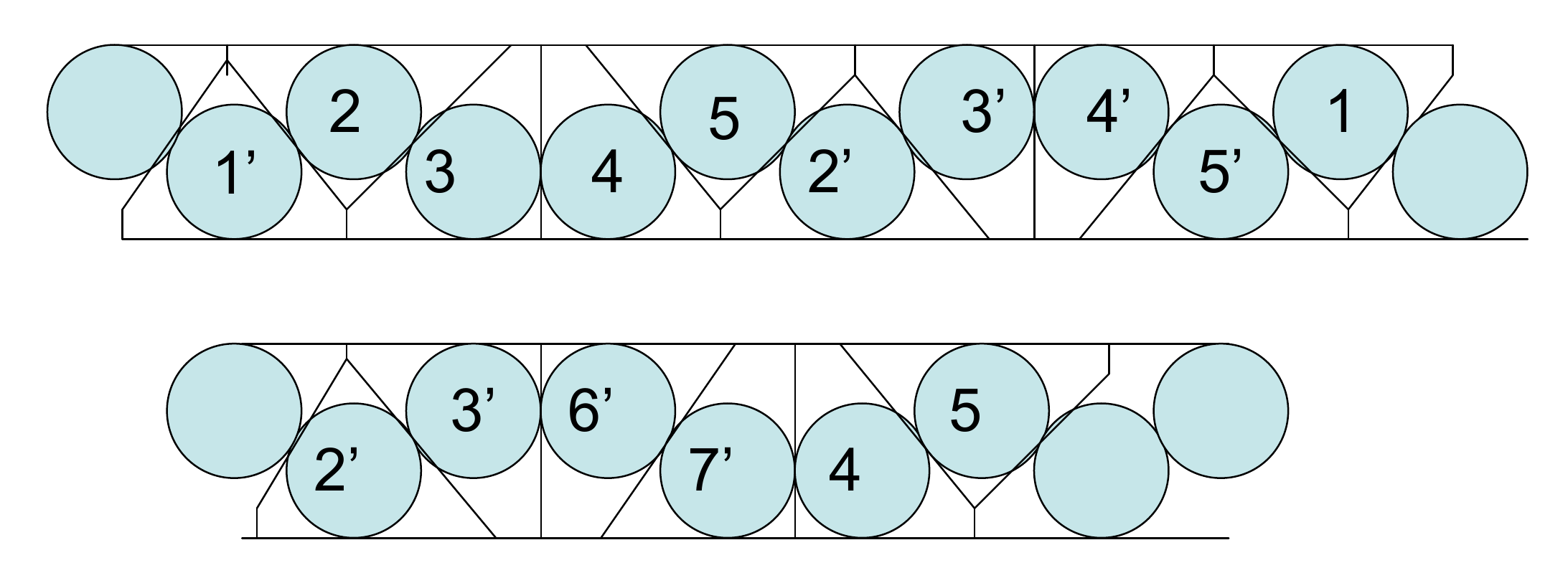}
 \label{tiles}
\end{figure}
%%%%%%%%%%%%%%%%%%%%%%%%%%%%%%%%%

\newpage
\newpage
\newpage
\pagebreak[10]
\begin{figure}[b]
\centering
 \includegraphics[width=180mm]{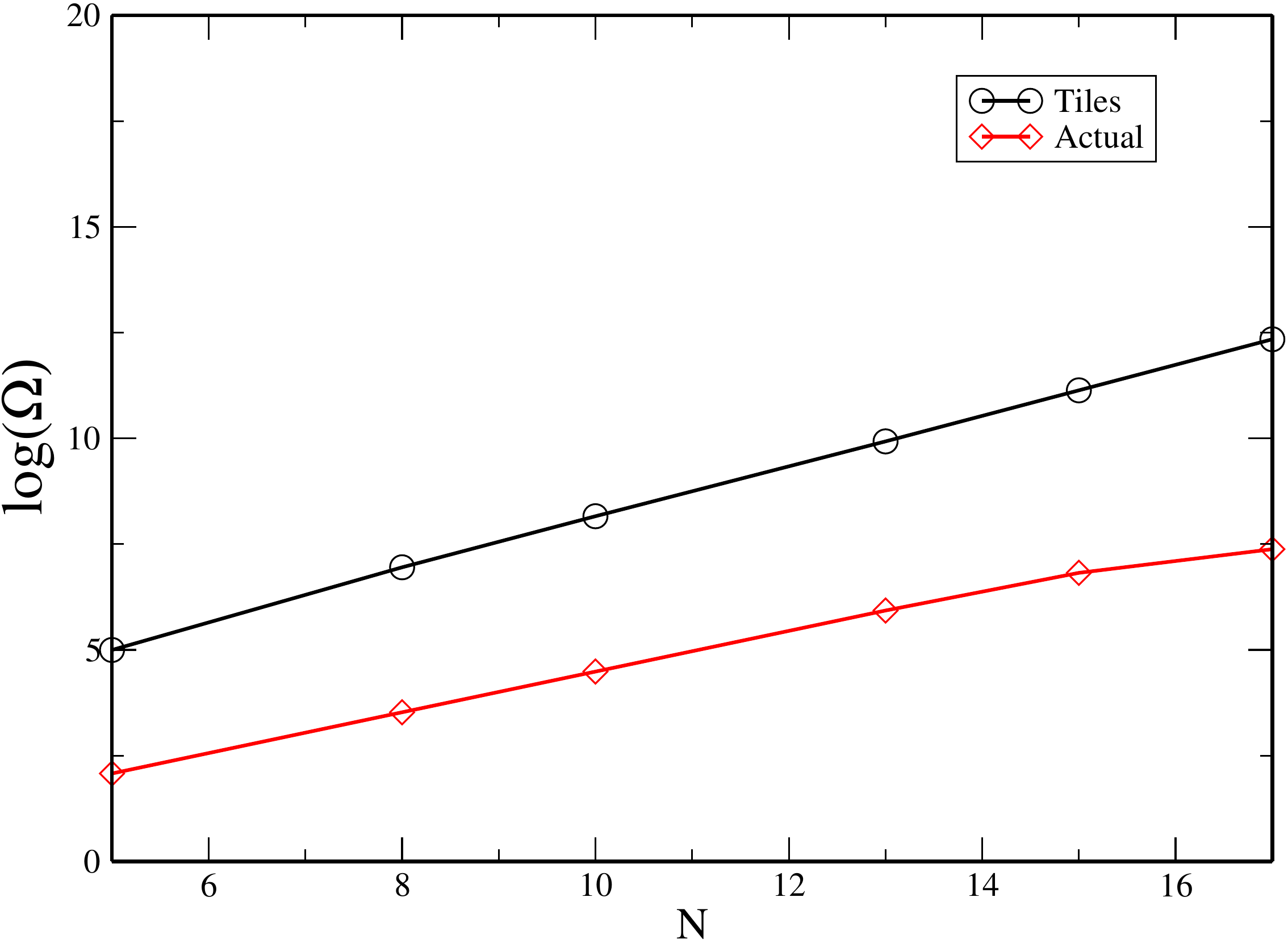}
 \label{tiles2}
\end{figure}

\newpage
\newpage
\ 

\end{document}